\documentclass[aps,pra,twocolumn,amssymb,amsfonts,floatfix,superscriptaddress,longbibliography,notitlepage ]{revtex4-2}
\usepackage[utf8]{inputenc}
\pdfoutput=1
\usepackage{color}
\usepackage{float}
\usepackage{xcolor}
\usepackage{mathtools,amsmath,amsthm,amsfonts,amssymb}
\usepackage{commath}
\usepackage{physics}
\usepackage{bm}
\usepackage[pdftex]{graphicx}
\usepackage[colorlinks=false]{hyperref}
\usepackage{enumitem}
 
\begin{document}
\DeclarePairedDelimiterX{\energyshellaverage}[2]{\langle}{\rangle_{\displaystyle #2}}{#1}
\DeclarePairedDelimiterX{\energyshellaveragelong}[3]{\langle}{\rangle_{#2\in\mathcal{M}(#3)}}{#1}

\title{Effective dimensions of infinite-dimensional Hilbert spaces: A phase-space approach}

\author{Sa\'ul Pilatowsky-Cameo}
\affiliation{Instituto de Ciencias Nucleares, Universidad Nacional Aut\'onoma de M\'exico, Apdo. Postal 70-543, C.P. 04510 CDMX, Mexico}
\affiliation{Center for Theoretical Physics, Massachusetts Institute of Technology, Cambridge, Massachusetts 02139, USA
}
\author{David Villase\~nor}
\affiliation{Instituto de Ciencias Nucleares, Universidad Nacional Aut\'onoma de M\'exico, Apdo. Postal 70-543, C.P. 04510 CDMX, Mexico}
\author{Miguel A. Bastarrachea-Magnani}
\affiliation{Departamento de F\'isica, Universidad Aut\'onoma Metropolitana-Iztapalapa, San Rafael Atlixco 186, C.P. 09340 CDMX, Mexico}
\author{Sergio Lerma-Hern\'andez}
\affiliation{Facultad de F\'isica, Universidad Veracruzana, Circuito Aguirre Beltr\'an s/n,  C.P. 91000 Xalapa, Veracruz, Mexico}
\author{Jorge G. Hirsch}
\affiliation{Instituto de Ciencias Nucleares, Universidad Nacional Aut\'onoma de M\'exico, Apdo. Postal 70-543, C.P. 04510  CDMX, Mexico}


\begin{abstract}

By employing Husimi quasiprobability distributions, we show that a bounded portion of an unbounded phase space induces a finite effective dimension in  an infinite dimensional Hilbert space. We compare our general expressions with numerical results for the spin-boson Dicke model in the chaotic energy regime, restricting its unbounded four-dimensional phase space to a classically chaotic energy shell. This effective dimension can be employed to characterize quantum phenomena in infinite dimensional systems, such as localization and scarring.

\end{abstract}

\maketitle

\section{Introduction}

If the phase space associated with a quantum system has finite volume, then the Hilbert space must be finite-dimensional. A bounded phase space can only ``accommodate a finite number of Planck size cells, and therefore a finite number of orthogonal quantum states''~\cite{Rovelli2015}. 

Quasiprobability distributions, such as the Husimi function~\cite{Husimi1940,Takahashi1986,Takahashi1986B,Takahashi1988,Takahashi1989,Takahashi1993}, may be used to study the distribution of a quantum state in the phase space. 
Even when the phase space is unbounded, normalization prevents a state from extending beyond a bounded portion. Thus, relevant information about the state may be extracted from the distribution of its Husimi function within a fixed bounded portion of the phase space.

The Dicke model~\cite{Dicke1954} is a fundamental model of quantum optics that has become a paradigm of the study of quantum chaos~\cite{Muller1991,Deaguiar1991,Deaguiar1992,Finney1996,Lambert2004,Lambert2005,Bastarrachea2014b,Chavez2019,Pilatowsky2020}. It has an unbounded four-dimensional phase space, but the energy shells in the classical limit are bounded, such that by averaging the moments of the Husimi function over them it is possible to define relative phase-space occupation measures~\cite{Wang2020,Pilatowsky2021NatCommun,Villasenor2021,Lozej2022}, which gauge the spreading of a quantum state within a single classical energy shell. In Ref.~\cite{Pilatowsky2021Identification} it is shown that these measures may be used to detect quantum scars~\cite{Heller1984,Villasenor2020,Pilatowsky2021}, and their maximum value for pure states can be derived from the average entropy of random states in a finite-dimensional Hilbert space~\cite{Jones1990}. The latter result is intriguing, given that the Dicke model is not finite-dimensional.

If a bounded phase space implies a finite-dimensional Hilbert space, does a bounded portion of an unbounded phase space generate a finite effective dimension in an infinite-dimensional Hilbert space? In this work, we explore this question and show that the answer is affirmative. We exhibit that it is possible to define an effective dimension associated with the classical energy shells of the Dicke model through averages of the Husimi function of random states. This explains why in Ref.~\cite{Pilatowsky2021Identification} the localization of random states within the classical energy shells is described by that of random states in a finite-dimensional Hilbert space: the energy shells induce a finite dimension within the infinite-dimensional Hilbert space.

The article is organized as follows. In Sec.~\ref{sec:DickeModel} we introduce the Dicke model, its classical limit and its associated phase space. We also describe a general expression to calculate averages within single energy shells in this phase space. In Sec.~\ref{sec:EffectiveDimension} we define an effective dimension for  classical energy shells, which relies on the Husimi function of random pure states. Next, in Sec.~\ref{sec:EffecExp}, we obtain analytical expressions for the effective dimension of classical energy shells, and we compare them to numerical results in the chaotic energy region of the Dicke model. In Sec.~\ref{sec:EffectiveDimensionPR} we contrast this effective dimension and the quantum participation ratio with each other using random states with a rectangular energy profile.  Finally, we present our conclusions in Sec.~\ref{sec:Conclusions}.

\section{Dicke Model}
\label{sec:DickeModel}

As a general model of spin-boson interaction, the Dicke model is widely used in physics, specifically in quantum optics, to describe atoms interacting with electromagnetic fields within a cavity~\cite{Dicke1954}. The most common picture of the model takes into account a set of $\mathcal{N}$ two-level atoms with excitation energy $\omega_{0}$ (using $\hbar = 1$), and a single-mode electromagnetic field with radiation frequency $\omega$. Their interaction is modulated by the atom-field coupling strength $\gamma$, whose critical value $\gamma_{c}=\sqrt{\omega\omega_{0}}/2$ divides two phases in the model. That is, the system develops a quantum phase transition going from a normal phase ($\gamma<\gamma_c$) to a superradiant phase ($\gamma>\gamma_c$)~\cite{Hepp1973a,Hepp1973b,Wang1973,Emary2003}.

The  Hamiltonian of the Dicke model is given by
\begin{align}
    \label{eqn:qua_hamiltonian}
    \hat{H}_{D} & = \hat{H}_{0} + \hat{H}_{\gamma}, \\
    \hat{H}_{0} & = \omega\hat{a}^{\dagger}\hat{a} + \omega_{0}\hat{J}_{z}, \\
    \hat{H}_{\gamma} & = \frac{\gamma}{\sqrt{\mathcal{N}}}(\hat{a}^{\dagger}+\hat{a})(\hat{J}_{+}+\hat{J}_{-}),
\end{align}
where $\hat{H}_{0}$ is the non-interacting Hamiltonian and $\hat{H}_{\gamma}$ includes the atom-field interaction. $\hat{a}^{\dagger}$ ($\hat{a}$) is the bosonic creation (annihilation) operator of the field mode which satisfies the Heisenberg-Weyl algebra, and $\hat{J}_{+}$ ($\hat{J}_{-}$) is the raising (lowering) collective pseudo-spin operator, defined by $\hat{J}_{\pm}=\hat{J}_{x}\pm i\hat{J}_{y}$. The collective pseudo-spin operators are defined by means of the Pauli matrices $\hat{\sigma}_{x,y,z}$ as $\hat{J}_{x,y,z}=(1/2)\sum_{k=1}^{\mathcal{N}}\hat{\sigma}_{x,y,z}^{k}$, and satisfy the SU(2) algebra.

The squared total pseudo-spin operator, $\hat{\textbf{J}}^{2}=\hat{J}_{x}^{2}+\hat{J}_{y}^{2}+\hat{J}_{z}^{2}$, has eigenvalues $j(j+1)$. These values correspond to different invariant atomic subspaces of the model. Here, we work with the totally symmetric subspace, defined by the maximum pseudo-spin value $j=\mathcal{N}/2$ that includes the ground state. Although the atomic sector is finite, the complete Hilbert space of the model is infinite due to the bosonic sector. However, wave functions can be computed to arbitrary numerical precision by appropriately truncating the bosonic sector.

The Dicke model has a rich combination of chaotic and regular behavior displayed as a function of the Hamiltonian parameters. In this work, we consider the resonant frequency case $\omega=\omega_0=1$, a coupling strength in the superradiant phase, $\gamma=2\gamma_c=1$, and we use rescaled energies $\epsilon=E/j$ to the system size $j=100$. For this set of Hamiltonian parameters, the classical dynamics is fully chaotic at  energies $\epsilon \geq -0.8$~\cite{Chavez2016}.

\subsection{Classical model and phase space}

The bosonic Glauber and the atomic Bloch coherent states, represented by the canonical variables $(q,p)$ and $(Q,P)$, respectively, are defined as
\begin{align}
    \label{eq:coherentstates}
    |q,p\rangle & =e^{-(j/4)\left(q^{2}+p^{2}\right)}e^{\left[\sqrt{j/2}\left(q+ip\right)\right]\hat{a}^{\dagger}}|0\rangle, \\
    |Q,P\rangle & =\Big(1-\frac{Q^2+P^2}{4}\Big)^{j}e^{\left[\left(Q+iP\right)/\sqrt{4-Q^2-P^2}\right]\hat{J}_{+}}|j,-j\rangle, \nonumber
\end{align}
with $|0\rangle$ the photon vacuum and $|j,-j\rangle$ the state with all the atoms in the ground state. By considering the tensor product of these coherent states $|\bm x\rangle=|q,p\rangle\otimes~|Q,P\rangle$, a classical Hamiltonian for the Dicke model can be obtained~\cite{Deaguiar1991,Deaguiar1992,Bastarrachea2014a,Bastarrachea2014b,Bastarrachea2015,Chavez2016,Villasenor2020}. Taking the expectation value of the quantum Hamiltonian $\hat{H}_{D}$ in these states and dividing by the size $j$ of the atomic sector, one gets
\begin{align}
    \label{eq:class_Hamiltonian}
    h_\text{cl}(\bm x) & =\frac{\langle\bm{x}|\hat{H}_{D}|\bm{x}\rangle}{j} = h_{0}(\bm x) + h_{\gamma}(\bm x), \\
    h_{0}(\bm x) & = \frac{\omega}{2}\big(q^{2}+p^{2}\big)+\frac{\omega_{0}}{2}\big(Q^2+P^2\big)-\omega_{0}, \\
    h_{\gamma}(\bm x) & =2\gamma qQ\sqrt{1-\frac{Q^2+P^2}{4}},
\end{align}
where $h_{0}(\bm x)$ represents the Hamiltonian of two harmonic oscillators and $h_{\gamma}(\bm x)$ is a non-linear coupling between them.

The phase space $\mathcal{M}$  of the classical Hamiltonian $h_\text{cl}(\bm x)$ is four-dimensional, with canonical variables $\bm x=(q,p;Q,P)$. While the bosonic variables $(q,p)$ can take any real value, the atomic variables $(Q,P)$ are bounded  by $Q^2 + P^2 \leq 4$. This phase space can be partitioned into a family of classical energy shells given by
\begin{equation}
    \mathcal{M}(\epsilon)=\{\bm x\in \mathcal{M} \,\mid\, h_\text{cl}(\bm x)=\epsilon\}, 
\end{equation}
where the rescaled classical energy $\epsilon=E/j$ defines an effective Planck constant $\hbar_{\text{eff}}=1/j$~\cite{Ribeiro2006}. The finite volume of the classical energy shells $\mathcal{M}(\epsilon)$ is given by 
\begin{equation}
    \label{eq:volumeCES}
    \int_{\mathcal{M}}\dif \bm x\,  \delta(h_\text{cl}(\bm x)-\epsilon) =(2\pi\hbar_{\text{eff}})^2\nu(\epsilon),
\end{equation}
where $\nu(\epsilon)$ is a semiclassical approximation to the quantum density of states (in units of $\epsilon^{-1}$) obtained with the Gutzwiller trace formula~\cite{Gutzwiller1971,Gutzwiller1990book} and whose analytical expression \footnote{See Ref. \cite{Bastarrachea2014a} for more details on $\nu(\epsilon)$, but mind the different units of $1/E$ instead of $1/\epsilon$} can be found in Ref.~\cite{Pilatowsky2021Identification}.

We calculate the average $ \energyshellaverage*{f}{\epsilon}$ of an arbitrary function $f(\bm x)$ 
in the bounded classical energy shells of the Dicke model $\mathcal{M}(\epsilon)$ by integrating $f(\bm x)$ with respect to the three-dimensional surface measure $\dif \bm x\, \delta(h_\text{cl}(\bm x) - \epsilon)$ and dividing the result by the total volume of the energy shell $(2\pi\hbar_{\text{eff}})^2\nu(\epsilon)$ [see Eq.~\eqref{eq:volumeCES}]~\cite{Pilatowsky2021Identification}
\begin{align}
    \label{eqn:averageCES}
    \energyshellaverage*{f}{\epsilon} & =\energyshellaveragelong*{f(\bm x)}{\bm x}{\epsilon} \nonumber \\
    & \equiv \frac{1}{(2\pi\hbar_{\text{eff}})^2\nu(\epsilon)}\int_{\mathcal{M}}\!\!\dif \bm x\,  \delta(h_\text{cl}(\bm x)-\epsilon) f(\bm x).
\end{align}

\section{Effective Dimension for Classical Energy Shells}
\label{sec:EffectiveDimension}

In a finite Hilbert space of dimension $N$, the average of the squared projection of an arbitrary state $\ket{\Psi}$ over all possible (normalized) states $\ket{\Phi}$, that is $\expval*{\abs*{\braket{\Psi}{\Phi}}^{2}}_{\ket{\Phi}}$, has been explicitly calculated in Ref.~\cite{Jones1990}. This average is given by
\begin{equation}
    \label{eq:formulascalingOfaverageAlpha1}
    \expval{\abs{\braket{\Psi}{\Phi}}^{2}}_{\ket{\Phi}} = \frac{\Gamma(N)}{\Gamma(N +1)} = \frac 1 N,
\end{equation}
where $\Gamma$ is the Gamma function and the identity $\Gamma(N+1) = N \, \Gamma(N)$ was used.

If, instead of averaging over the whole Hilbert space, we consider averages over normalized states in a vector subspace $W$  with dimension $N_{W}<N$, since  $|\Psi\rangle$ may have orthogonal components to the vector subspace,  we obtain
\begin{equation}
    \label{eq:inequality}
    \expval{\abs{\braket{\Psi}{\phi}}^{2}}^{-1}_{|\phi\rangle\in W} \geq N_{W}.
\end{equation}
The  equality holds if and only if $|\Psi\rangle$ belongs to the subspace $W$.  The inequality remains valid if we consider an additional average, but now over an ensemble $  \mathcal{S}$ of states $|\Psi\rangle$,
\begin{equation}
    \label{eq:inequality2}
    \left\langle\expval{\abs{\braket{\Psi}{\phi}}^{2}}_{|\phi\rangle\in W}\right\rangle_{|\Psi\rangle\in\mathcal{S}}^{-1} \geq N_{W}.
\end{equation}

\subsection{Dimensionality and Effective Dimension}
An arbitrary pure state $|\psi\rangle$ of the Dicke model can be represented in phase space employing the Husimi quasi-probability distribution~\cite{Husimi1940}, $\mathcal{Q}_{\psi}(\bm x)=\abs{\braket{\psi}{\bm x}}^{2}\geq 0$. The average of the Husimi function over a classical energy shell $\mathcal{M}(\epsilon)$
\begin{equation}
    \label{eqn:HusimiMoments}
    \energyshellaverage{\mathcal{Q}_{\psi}}{\epsilon}\equiv\energyshellaveragelong*{\abs{\braket{\psi}{\bm x}}^{2}}{\bm x}{\epsilon},
\end{equation} 
can be obtained with Eq.~\eqref{eqn:averageCES} using $f(\bm x) = \mathcal{Q}_{\psi}(\bm x) = \abs{\braket{\psi}{\bm x}}^{2}$.  Equation~\eqref{eqn:HusimiMoments} is similar to the average in Eq.~\eqref{eq:inequality}, taking $W$ to be the set of coherent states $|\bm x\rangle$ with $\bm x \in \mathcal{M}(\epsilon)$. However, this set is not a vector subspace, and, consequently, there may be states in the Hilbert space that are strongly correlated with all the members of the set. In order to eliminate these  correlations, we  consider an ensemble $\mathcal{S}$ of random pure states $|\psi_{\text{R}}\rangle$. 

Because the Dicke model has an infinite spectrum, we consider random states whose energy components  follow 
a given energy profile $\rho_{\text{R}}(\epsilon)\geq 0$ such that $\int \dd \epsilon\, \rho_{\text{R}}(\epsilon)=1$. The components of the random states will be weighted by this energy profile, such that the resulting random states are contained within it. The shape of the energy profile in principle could be arbitrary. In Sec.~\ref{sec:EffecExp} we study in detail the cases where $\rho_{\text{R}}$ is a rectangular and a Gaussian    profile. We center $\rho_{\text{R}}$ at a fixed energy and use the average value over different random states with the same energy profile to define the effective dimension of the energy shell, $\mathcal{D}_{\text{eff}}(\epsilon)$. This is done by considering an inequality similar to Eq.~\eqref{eq:inequality2}, as follows. 

We construct random states $|\psi_{\text{R}}\rangle$ in the energy eigenbasis, 
\begin{equation}
    \label{eq:randef}
    |\psi_{\text{R}}\rangle=\sum_k c_k |\varphi_k\rangle,
\end{equation}
where $\hat{H}_{D}|\varphi_k\rangle= E_k|\varphi_k\rangle$.
The components $c_k$ are complex numbers with random phases and magnitudes   given by~\cite{Villasenor2020}
\begin{equation}
    \label{eq:compon}
    |c_k|^2= \frac{r_k \, \rho_{\text{R}}(\epsilon_k)}{ M \,\nu(\epsilon_k)},
\end{equation}
where $r_k$ are positive random numbers from an arbitrary distribution whose first momentum is $\langle r\rangle$. As shown in App.~\ref{app:EffectiveWidthDerivation}, the results we present below are independent of the exact form of the distribution of $r_k$.  $M$ is a normalization constant that is approximately given by $M\approx \langle r\rangle$ (see App.~\ref{app:EffectiveWidthDerivation} for details). The density of states $\nu(\epsilon)$ in the denominator ensures that the members of the ensemble $\mathcal{S}$ have the chosen energy profile $\rho_{\text{R}}(\epsilon)$.

Now, we consider Eq.~\eqref{eqn:HusimiMoments} and take the inverse of its average over the ensemble of random states  $|\psi_{\text{R}}\rangle$ with energy profile $\rho_{\text{R}}$,
\begin{align}
    \label{eq:dimendef}
    D(\epsilon,\rho_{\text{R}})\equiv \left \langle \energyshellaverage*{\mathcal{Q}_{\psi_{\text{R}}}}{\epsilon} \right\rangle_{\psi_{\text{R}}}^{-1}.
\end{align}  
We call $D(\epsilon,\rho_{\text{R}})$ the dimensionality of the ensemble of random states over the energy shell $\mathcal{M}(\epsilon)$.  

Next, inspired by Eq.~\eqref{eq:inequality2},  we define the effective dimension $\mathcal{D}_\text{eff}(\epsilon)$ of the classical energy shell $ \mathcal{M}(\epsilon)$ 
by minimizing the dimensionality $D(\epsilon,\rho_{\text{R}})$ with respect to arbitrary energy profiles  
\begin{equation}
    \label{eq:defNeffective}
    \mathcal{D}_\text{eff}(\epsilon)\equiv \min_{\rho_{\text{R}}}\left[ D(\epsilon, \rho_{\text{R}})
    \right].
\end{equation}
It immediately follows that 
\begin{equation}
    \label{eq:NeffDefinition}
    D(\epsilon,\rho_{\text{R}})   
    \equiv \left\langle\energyshellaveragelong*{\abs{\braket{\psi_{\text{R}}}{\bm x}}^{2}}{\bm x}{\epsilon} \right\rangle_{\psi_{\text{R}}}^{-1} \geq  \mathcal{D}_\text{eff}(\epsilon),
\end{equation}
which has the same form as Eq.~\eqref{eq:inequality2}. We emphasize that the set of coherent states $|\bm{x}\rangle$ in the classical energy shell $\mathcal{M}(\epsilon)$ is not a vector subspace, but Eq.~\eqref{eq:defNeffective} allows to measure the number of orthonormal states available for this set~\cite{Perelomov1971, BARGMANN1971221}. 

Moreover, we can define the dimensionality of a given eigenstate $|\varphi_k\rangle$, with eigenenergy $\epsilon_k=E_k/j$, over an energy shell $\mathcal{M}(\epsilon)$, by considering the particular energy profile $\rho_k(\epsilon)=\delta(\epsilon-\epsilon_k)$. This  gives the inverse of the average of the Husimi function of eigenstate  $|\varphi_k\rangle$ over the classical energy shell $\mathcal{M}(\epsilon)$,
\begin{equation}
    \label{eqn:EigenDimensionality}
    D(\epsilon,\varphi_k) \equiv D(\epsilon, \rho_k) = \energyshellaverage{\mathcal{Q}_{\varphi_k}}{\epsilon}^{-1},    
\end{equation}
where the average is calculated with Eq.~\eqref{eqn:HusimiMoments} as
\begin{equation}
    \label{eqn:EigenAverage}
    \energyshellaverage
    {\mathcal{Q}_{\varphi_k}}{\epsilon}=\energyshellaveragelong*{\abs{\braket{\varphi_{k}}{\bm x}}^{2}}{\bm x}{\epsilon}.
\end{equation}

In the following section, we provide analytical and  numerical evidences which show that the minimum value of the dimensionality $D(\epsilon,\rho_{\text{R}})$ is attained for energy profiles $\rho_{\text{R}}$ centered at $\epsilon=\int d\epsilon' \rho_{\text{R}}(\epsilon') \epsilon'$,  with an energy standard deviation 
$\sigma_R=\sqrt{\int d\epsilon' \rho_{\text{R}}(\epsilon') (\epsilon'-\epsilon)^2 }$
which is much smaller than the energy standard deviation of the coherent states given by Eq. \eqref{eq:coherentstates}. An analytical expression for this minimum is also provided and  shown to be approximately  equal to the dimensionality of the eigenstate $\epsilon_{k}$ closest to $\epsilon$,
\begin{equation}
    \mathcal{D}_\text{eff}(\epsilon)\approx D(\epsilon_k\approx\epsilon,\varphi_k) \equiv
    \left\langle\mathcal{Q}_{\varphi_k}\right\rangle_{\epsilon_k\approx\epsilon}^{-1}.
\end{equation}

\begin{figure*}
    \centering
    \includegraphics[width=0.95\textwidth]{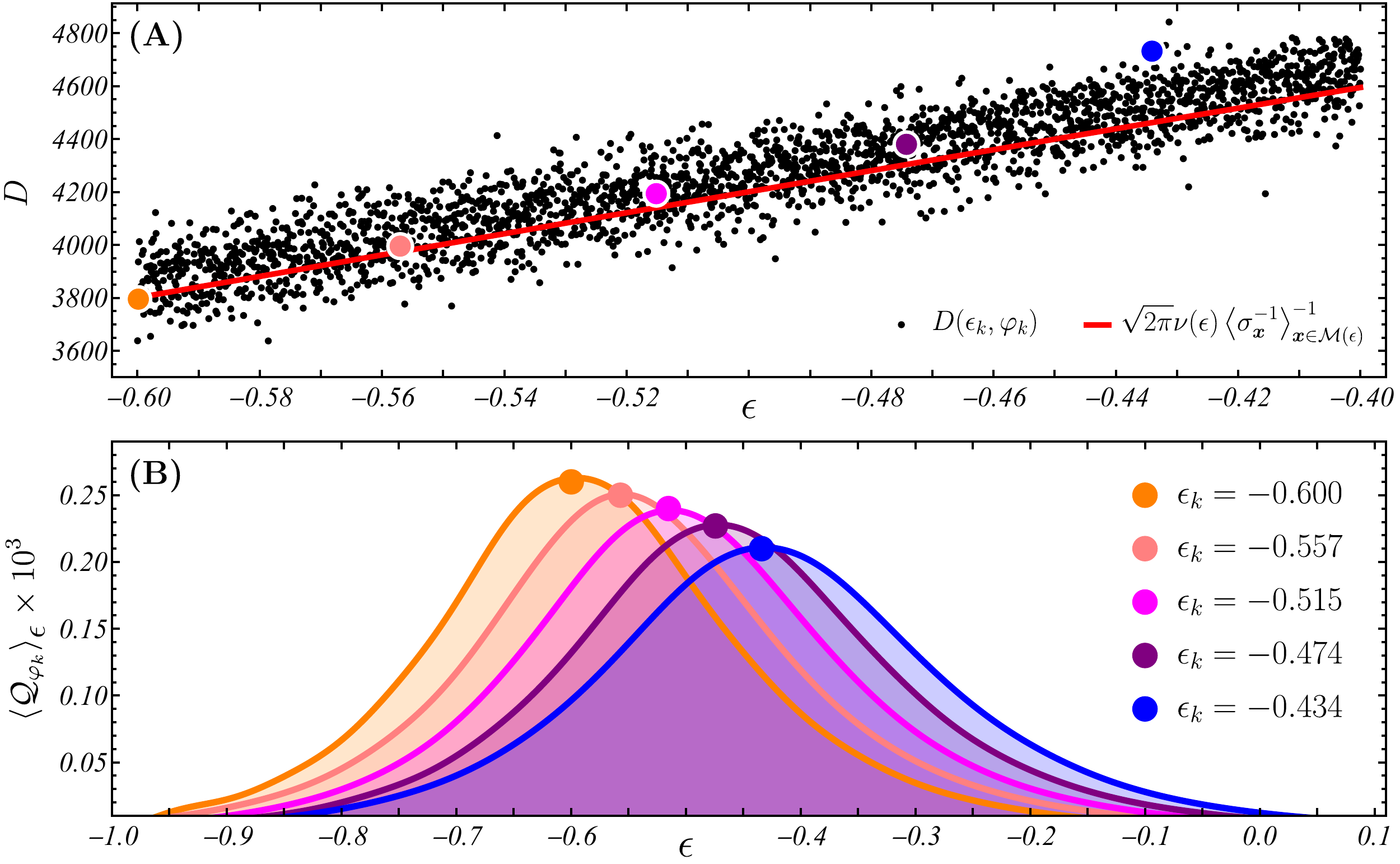}
    \caption{Panel (A): Dimensionality $D(\epsilon_k,\varphi_k)=\energyshellaverage
    {\mathcal{Q}_{\varphi_k}}{\epsilon_k}^{-1}$ (black circles) [see Eq.~\eqref{eqn:EigenDimensionality}] for all eigenstates $|\varphi_k\rangle$ with eigenenergies contained in the energy interval $\epsilon_k\in[-0.6,-0.4]$ and analytical approximation (red solid line) given by Eq.~\eqref{eq:eigenstateseffdim}. Panel (B): Energy profiles of the Husimi function $\energyshellaverage
    {\mathcal{Q}_{\varphi_k}}{\epsilon}$ (colored solid curves) [see Eq.~\eqref{eqn:EigenAverage}] as a function of the energy $\epsilon$ for some eigenstates contained in the same energy interval $\epsilon_k\in[-0.6,-0.4]$. The selected eigenstates are specified by the energy spectrum indices $k=7960,8460,8960,9460,9960$ and their eigenenergies are indicated by the colored solid dots, which are also shown in panel (A). The system size in both panels (A) and (B) is $j=100$.}
    \label{fig01}
\end{figure*}


\begin{figure}
    \centering
    \includegraphics[width=0.95\columnwidth]{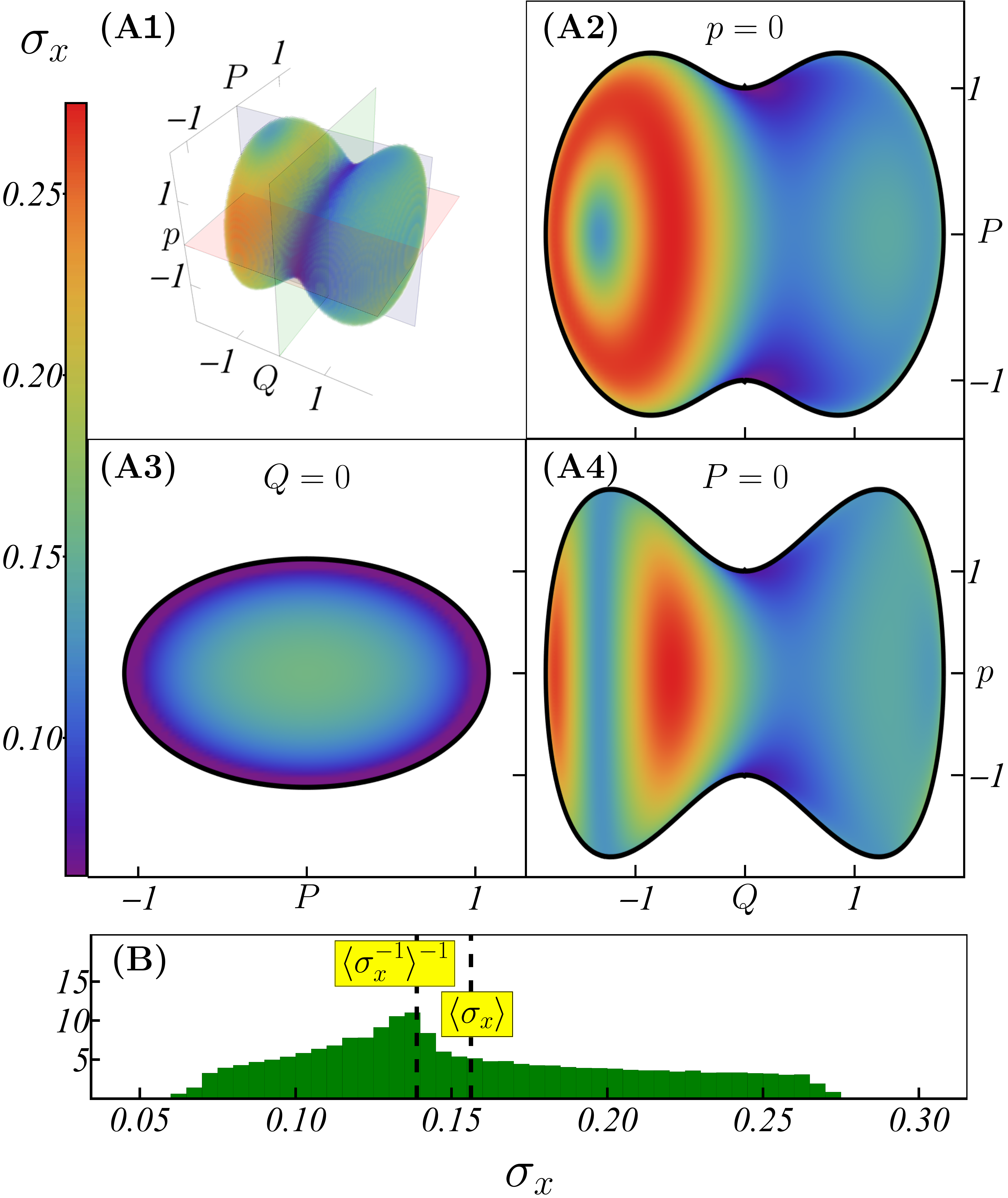}
    \caption{Panels (A1)-(A4): 3D distribution (A1) of coherent-state energy standard deviation $\sigma_{\boldsymbol{x}}$ [see Eq.~\eqref{eq:CSwidth}] in the classical energy shell at $\epsilon=-0.5$, and its values along  orthogonal planes $Q-P$ (A2), $Q-p$ (A3), and $P-p$ (A4). In panel (A1) the color planes represent each orthogonal plane: red (A2), green (A3), and blue (A4). The left color scale depicts the distribution of the energy standard deviations from the smaller (purple) to the larger (red). Panel (B): Statistical distribution of coherent-state energy standard deviations $\sigma_{\boldsymbol{x}}$ displayed in panel (A1). Vertical black dashed lines depict the harmonic $\langle \sigma_{\boldsymbol{x}}^{-1} \rangle^{-1}_{\bm x \in \mathcal{M}(\epsilon)}=0.1389$ and the standard $\langle \sigma_{\boldsymbol{x}} \rangle_{\bm x \in \mathcal{M}(\epsilon)}=0.1563$ mean of these standard deviations, respectively (see the yellow boxes). The system size in all panels (A1)-(A4) and (B) is $j=100$.}
    \label{fig02}
\end{figure}


\section{Finding the Minimum of the Dimensionality}
\label{sec:EffecExp}

In this section we discuss the process to find the minimum value of the dimensionality $D(\epsilon,\rho_{\text{R}})$.
By substituting the random states of Eq.~\eqref{eq:randef} in Eq.~\eqref{eq:dimendef}, we obtain
\begin{equation}
    \label{eq:dimenwithcomp}
    D(\epsilon,\rho_\text{R})
    \approx
    \left( \sum_k \langle \abs{c_k}^2\rangle_{\psi_{\text{R}}} 
    \energyshellaverage{\mathcal{Q}_{\varphi_k}}{\epsilon}
    \right)^{-1}, 
\end{equation}
where, as it is shown in App.~\ref{app:EffectiveWidthDerivation}, we have used that the phases of the components of $|\psi_{\text{R}}\rangle$ are randomly distributed. 
The dimensionality  can be further simplified by performing the ensemble average of the squared magnitude of the components \eqref{eq:compon}. This calculation, whose details  are presented in App.~\ref{app:EffectiveWidthDerivation}, leads to 
\begin{align}
    \label{eq:generaleffdim}
    D(\epsilon,\rho_\text{R})
    & \approx \left( \sum_k \frac{\rho_{\text{R}}(\epsilon_k)}{\nu(\epsilon_k)  
    } \energyshellaverage{\mathcal{Q}_{\varphi_k}}{\epsilon}
    \right)^{-1}. 
\end{align}
To find the minimum value of the dimensionality, it is necessary  to obtain an expression for the average of the Husimi function of eigenstates over arbitrary energy shells $\energyshellaverage{\mathcal{Q}_{\varphi_k}}{\epsilon}$.  Such an expression can be obtained by using generic properties of the coherent states $|\bm{x}\rangle$, as explained in the following. 

\subsection{Average of Eigenstates over Classical Energy Shells}

We turn our attention to the average of the  Husimi function of an eigenstate  with eigenenergy $\epsilon_k=E_k/j$ over a classical energy shell $\mathcal{M}(\epsilon)$. 
%
By assuming that both the energy of the classical shell $\epsilon$ and the eigenenergy $\epsilon_k$ are far enough from the ground state energy, it can be shown that (see App.~\ref{app:AveHusEigen} for details) 
\begin{equation}
    \label{eq:eigenstateseffdimEp}
    \energyshellaverage
    {\mathcal{Q}_{\varphi_k}}{\epsilon} \approx \frac{1}{\sqrt{2\pi}\nu(\epsilon_k)}\energyshellaveragelong*{\frac{\exp(\frac{-(\epsilon_k-\epsilon)^2}{2\sigma_{\bm x}^2})}{\sigma_{\bm x}}}{\bm x}{\epsilon},
\end{equation}
where $\sigma_{\bm x}$ is the energy standard deviation of the coherent state $|\bm x\rangle$, 
\begin{equation}
    \label{eq:CSwidth}
    \sigma_{\bm x}=\frac{1}{j}\sqrt{\langle\bm{x}|\hat{H}_{D}^{2}|\bm{x}\rangle-\langle\bm{x}|\hat{H}_{D}|\bm{x}\rangle^{2}},
\end{equation}
which can be calculated  analytically 
in the Dicke model~\cite{Schliemann2015,Lerma2018} (the analytical expression is shown in App.~\ref{app:EnergyWidthDerivation}).

As the energy of the classical shell  $\epsilon$ moves away from  the corresponding eigenenergy $\epsilon_k$, the value given by  Eq.~\eqref{eq:eigenstateseffdimEp} decays exponentially. In contrast, if we take the average of the eigenstate Husimi function over  the classical energy shell of its eigenenergy,  $\epsilon=\epsilon_k$, then the shell average of the eigenstate Husimi function attains its maximum value
, which is  given by  Eq.~\eqref{eq:eigenstateseffdimEp} as
\begin{equation}
    \label{eq:eigenstateseffdim}
    \energyshellaverage
    {\mathcal{Q}_{\varphi_k}}{\epsilon_k}
    = \frac{1}{D(\epsilon_{k},\varphi_{k})} \approx \frac{1}{\sqrt{2\pi}\nu(\epsilon_k)\overline{\sigma}_{\text{c}}(\epsilon_k)},
\end{equation}
where $\overline{\sigma}_{\text{c}}(\epsilon_k)$ is 
the harmonic mean of the energy standard deviations $\sigma_{\bm x}$ of all coherent states contained in the classical energy shell $\mathcal{M}(\epsilon_k)$, that is
\begin{equation}
    \label{eq:averagesigma}
    \overline{\sigma}_{\text{c}}(\epsilon_k) \equiv \energyshellaveragelong*{\sigma_{\bm x}^{-1}}{\bm x}{\epsilon_k}^{-1},
\end{equation}
which is the inverse of the mean of the inverse elements, in contrast to the standard mean defined as $\energyshellaveragelong*{\sigma_{\bm x}}{\bm x}{\epsilon_k}$.

In Fig.~\ref{fig01}~(A) the black dots mark 
$D(\epsilon_k,\varphi_k)=\energyshellaverage{\mathcal{Q}_{\varphi_k}}{\epsilon_k}^{-1}$, 
    for all eigenstates of the Dicke model inside the chaotic energy interval $\epsilon_k\in [-0.6,-0.4]$. The red line plots the analytical approximation given by Eq.~\eqref{eq:eigenstateseffdim}. Note that this approximation captures the overall tendency. Figure~\ref{fig01}~(B) shows the averaged Husimi function $\energyshellaverage
    {\mathcal{Q}_{\varphi_k}}{\epsilon}$ as a function of the energy $\epsilon$ [see Eq.~\eqref{eq:eigenstateseffdimEp}] of some eigenstates $|\varphi_k\rangle$ contained in the same energy interval $\epsilon_k\in [-0.6,-0.4]$. All the selected eigenstates show an averaged Husimi function with a Gaussian shape. 
    In addition, this figure shows that indeed the maximum of $\energyshellaverage
    {\mathcal{Q}_{\varphi_k}}{\epsilon}$ is attained at $\epsilon \approx \epsilon_k$.

As mentioned above, the coherent-state energy standard deviations $\sigma_{\boldsymbol{x}}$ centered on the classical energy shell $\mathcal{M}(\epsilon)$ 
play a fundamental role in the building of the dimensionality $D(\epsilon,\rho_{\text{R}})$ of the ensemble of random states $\mathcal{S}$. Their values were evaluated over the classical energy shell at $\epsilon=-0.5$, which is at the center of the chaotic energy interval $\epsilon \in [-0.6,-0.4]$. In Fig.~\ref{fig02}~(A1) we show the three-dimensional (3D) distribution of energy standard deviations $\sigma_{\boldsymbol{x}}$ for all coherent states in the classical energy shell $\boldsymbol{x} \in \mathcal{M}(\epsilon=-0.5)$, as well as 
their values along three orthogonal planes $Q$-$P$ ($p=0$), $P$-$p$ ($Q=0$), and  $Q$-$p$ ($P=0$) in Figs.~\ref{fig02}~(A2)-(A4), which show more clearly the behavior of such energy standard deviations $\sigma_{\boldsymbol{x}}$ within the 3D distribution. It can be seen that the widest states are concentrated in the region with $Q<0$, while the thinnest ones are concentrated along the external ring in $Q=0$ [see Fig.~\ref{fig02}~(A3)]. In Fig.~\ref{fig02}~(B) we show the distribution of the energy standard deviations $\sigma_{\boldsymbol{x}}$ for the complete set of coherent states contained in the classical energy shell $\boldsymbol{x} \in \mathcal{M}(\epsilon=-0.5)$. This distribution concentrates around the harmonic mean $\energyshellaveragelong*{\sigma_{\bm x}^{-1}}{\bm x}{\epsilon}^{-1}=0.1389$, with a clear asymmetry: the number of narrow states is almost linear with the standard deviation, and almost constant for the wider states, which are less overall. In the same figure we plot the standard mean $\energyshellaveragelong*{\sigma_{\bm x}^{}}{\bm x}{\epsilon}=0.1563$. We see that these two mean values are similar, but do not  coincide exactly.

One important result in this work is the fact that the harmonic mean $\energyshellaveragelong*{\sigma_{\bm x}^{-1}}{\bm x}{\epsilon}^{-1}$, not the standard one $\energyshellaveragelong*{\sigma_{\bm x}^{}}{\bm x}{\epsilon}$, is the one that best approximates the mean value of the Husimi function for eigenstates  $\energyshellaverage{\mathcal{Q}_{\varphi_k}}{\epsilon_k}$, as shown in  Eq.~\eqref{eq:eigenstateseffdim}.

\subsection{Analytical Expression of the Effective Dimension}

Having analyzed the classical energy shell averages of eigenstates, we can now discuss the minimum of the dimensionality for random states $D(\epsilon,\rho_{\text{R}})$ and determine the effective dimension of the  classical energy shell $\mathcal{M}(\epsilon)$.

Given that $\energyshellaverage{\mathcal{Q}_{\varphi_k}}{\epsilon}$ 
decays exponentially as a function of the binomial $(\epsilon-\epsilon_k)^2$ [see Eq.~\eqref{eq:eigenstateseffdimEp}], the minimum of the dimensionality  is obtained for energy profiles centered at $\epsilon$.
From Eq.~\eqref{eq:dimenwithcomp}, it is clear that the dimensionality is minimized when the energy profile $\rho_{\text{R}}$ is concentrated on the classical energy shell, having an energy standard deviation 
$\sigma_{\text{R}}= \sqrt{\int d\epsilon' \rho_\text{R}(\epsilon')(\epsilon'-\epsilon)^2 }\sim 0 $.
For such profiles, only the eigenstates with energy close to the classical energy shell contribute. So
\begin{equation}
   \langle |c_k|^2 \rangle_{\psi_{\text{R}}}\sim \left\{\begin{array}{cr}
    \not=0\ \ \ \ &\text{for $\epsilon_k \approx \epsilon $}\\
    0 \ \ \ & \text{otherwise}
    \end{array}\right., 
\end{equation}
and it becomes evident that $D(\epsilon,\rho_{\text{R}})$ is bounded from below by $\energyshellaverage{\mathcal{Q}_{\varphi_k}}{\epsilon_k}^{-1}
\approx \sqrt{2\pi}\nu(\epsilon_k)\overline{\sigma}_{\text{c}}(\epsilon_k)$ [see Eq.~\eqref{eq:eigenstateseffdim}], allowing, thus, to determine the effective dimension of the classical energy shell as
\begin{equation}
    \label{eq:effDim}
    \mathcal{D}_\text{eff}(\epsilon)=    \sqrt{2\pi}\nu(\epsilon)\overline{\sigma}_{\text{c}}(\epsilon).
\end{equation}    
This effective dimension is plotted as a solid line in Fig.~\ref{fig01}~(A). The relatively small dispersion in the dots depicting $D(\epsilon_{k},\varphi_{k})$ is caused by the variance of the coherent-state energy standard deviations $\sigma_{\bm x}$ inside of the classical energy shell.

The order of magnitude of the effective dimension obtained in Fig.~\ref{fig01}~(A) can be understood by simple arguments: From the semiclassical analysis, each of the two degrees of freedom causes a scaling  $(2\pi\hbar_{\text{eff}})^{-1}=j$ in the density of states $\nu(\epsilon)$, which, consequently, scales proportional to $j^2$  in agreement with  Eq.~\eqref{eq:volumeCES}. Since we project the  coherent states into the classical energy shells, which have half a degree of freedom, $\sigma_{\bm x}$ and $\overline{\sigma}_{\text{c}}$ scale with $j^{-1/2}$, and then the effective dimension must scale as $j^2/j^{1/2}=j^{3/2}$. For the value of  $j$ we use in Fig.~\ref{fig01}~(A), $j=100$, this yields an order of magnitude of $10^3$, which is indeed what we obtain in the figure. 

Note that the density of states $\nu(\epsilon)$ is precisely the three-dimensional volume of the energy shell divided by the four-dimensional volume of a Planck cell [see Eq.~\eqref{eq:volumeCES}]. Consequently, Eq.~\eqref{eq:effDim} says that there are a total of $\mathcal{D}_\text{eff}(\epsilon)$ Planck cells contained in the four-dimensional phase-space region bounded by $\mathcal{M}(\epsilon - \Delta\epsilon)$ and $\mathcal{M}(\epsilon/2 + \Delta\epsilon/2)$, where $2\Delta\epsilon\sim \sqrt{2\pi}\overline{\sigma}_{\text{c}}(\epsilon)$.

\subsection{Random states with rectangular and Gaussian energy Profiles}

To illustrate that the minimum of the dimensionality $D(\epsilon,\rho_{\text{R}})$ is correctly described by Eq.~\eqref{eq:effDim},  we consider two particular ensembles of random pure states: first, one with a rectangular energy profile $\hat{\rho}_{\text{R}}^{r}$ and then a second with a Gaussian energy profile $\hat{\rho}_{\text{R}}^{g}$.

The  average of the components of an ensemble of states $|\psi_{\text{R}}^{r}\rangle$  with a rectangular energy profile $\rho_{\text{R}}^r$ is given by 
\begin{equation}
    \label{eq:rectangularprofile}
    \langle |c_{k}^{r}|^{2}\rangle_{\psi_{\text{R}}^{r}} = \left\{\begin{array}{ll}
        \frac{1}{2\sqrt{3}\nu(\epsilon_{k})\sigma_{r}} & \text{if } \epsilon_{k}\in[\epsilon-\sqrt{3}\sigma_{r},\epsilon+\sqrt{3}\sigma_{r}] \\
        0 & \text{otherwise} 
    \end{array}\right.,
\end{equation}
where $\epsilon$ is the center of the profile and  $\sigma_r$ is its energy standard deviation.
The number of eigenstates $|\varphi_{k}\rangle$ contained within this rectangular energy window is given approximately by $2\sqrt{3}\nu(\epsilon)\sigma_{r}$, where we have evaluated the density of states in the center of the distribution, $\nu(\epsilon_{k}=\epsilon)$. It is straightforward to show that, for this particular energy profile,   Eq.~\eqref{eq:generaleffdim}
 leads to  (see App.~\ref{app:EffectiveWidthDerivationRectangular} for the derivation)
\begin{equation}
    \label{eq:neffrectangularprofile}
    D(\epsilon,\rho_{\text{R}}^{r}) 
    = 2\sqrt{3}\nu(\epsilon)\sigma_{r}\left\langle \text{erf}\left(\sqrt{\frac{3}{2}}\frac{\sigma_{r}}{\sigma_{\bm{x}}}\right) \right\rangle^{-1}_{\bm{x}\in\mathcal{M}(\epsilon)},
\end{equation}
where $\text{erf}(x)=\frac{2}{\sqrt{\pi}}\int_0^x \dd{t}e^{-t^2}$ is the error function. Equation \eqref{eq:neffrectangularprofile}  has the following limiting behaviors
\begin{equation}
    \label{eq:recLim}
    D(\epsilon,\rho_\text{R}^{r})
    \approx \nu(\epsilon) \begin{cases} 
    2\sqrt{3}\,\sigma_{r}
    & \text{if } \sigma_{r} \gg 
    \overline{\sigma}_{\text{c}}(\epsilon)\\
 \sqrt{2\pi}\,\overline{\sigma}_{\text{c}}(\epsilon) & \text{if } \sigma_{r} \ll 
    \overline{\sigma}_{\text{c}}(\epsilon)
    \end{cases} .
\end{equation}

Now, for states $|\psi_{\text{R}}^{g}\rangle$ coming from a  random ensemble with a  normalized Gaussian energy profile $\rho_{\text{R}}^g$  centered in the classical energy shell at $\mathcal{M}(\epsilon)$, the average amplitudes are given by
\begin{equation}
    \label{eq:gaussianprofile}
    \langle|c_{k}^{g}|^{2}\rangle_{\psi_{\text{R}}^{g}} = \frac{\text{exp}\left(\frac{-(\epsilon_{k}-\epsilon)^{2}}{2\sigma_{g}^{2}}\right)}{\sqrt{2\pi}\nu(\epsilon_{k})\sigma_{g}},
\end{equation}
where $\sigma_{g}$ defines the energy standard deviation of the Gaussian profile. For this case, Eq.~\eqref{eq:generaleffdim} yields the following result for the dimensionality  (see App.~\ref{app:EffectiveWidthDerivationGauss} for the derivation) 
\begin{equation}
    \label{eq:neffgaussianprofile}
    D(\epsilon,\rho_{\text{R}}^{g})
    = \sqrt{2\pi}\nu(\epsilon)\sigma_{g}\left\langle \left[ 1+\left(\frac{\sigma_{\bm{x}}}{\sigma_{g}}\right)^{2} \right]^{-\frac{1}{2}} \right\rangle^{-1}_{\bm{x}\in\mathcal{M}(\epsilon)},
\end{equation}
with the corresponding limiting behaviors
\begin{equation}
    \label{eq:gaussLim}
    D(\epsilon,\rho_\text{R}^{g}) 
    \approx 
    \sqrt{2\pi}\nu(\epsilon)
    \begin{cases} 
    \sigma_{g} & \text{if } \sigma_{g} \gg \overline{\sigma}_{\text{c}}(\epsilon) \\
    \overline{\sigma}_{\text{c}}(\epsilon) 
    & \text{if } \sigma_{g} \ll \overline{\sigma}_{\text{c}}(\epsilon) 
    \end{cases} . 
\end{equation}

Equations~\eqref{eq:neffrectangularprofile}~and~\eqref{eq:neffgaussianprofile} and their respective asymptotic limits~\eqref{eq:recLim}~and~\eqref{eq:gaussLim} show that the dimensionalities increase with $\sigma_{r}$ and $\sigma_{g}$, the energy standard deviations of each energy profiles. In general, the dimensionalities depend on the particular energy profile chosen for the ensemble. However, if the energy standard deviation of the profile is much less than the harmonic mean of the coherent-state energy standard deviations in the classical energy shell $\overline{\sigma}_{\text{c}}(\epsilon)$, the dimensionality becomes independent of the energy profile and is given by  the reciprocal of the average of the Husimi function of eigenstates over their respective energy shells near the classical shell at energy $\epsilon$ [see Eq.~\eqref{eq:eigenstateseffdim}]. This confirms that, indeed, the effective dimension of the classical energy shell is given by Eq.~\eqref{eq:effDim}.

\begin{figure}
    \centering
    \includegraphics[width=0.95\columnwidth]{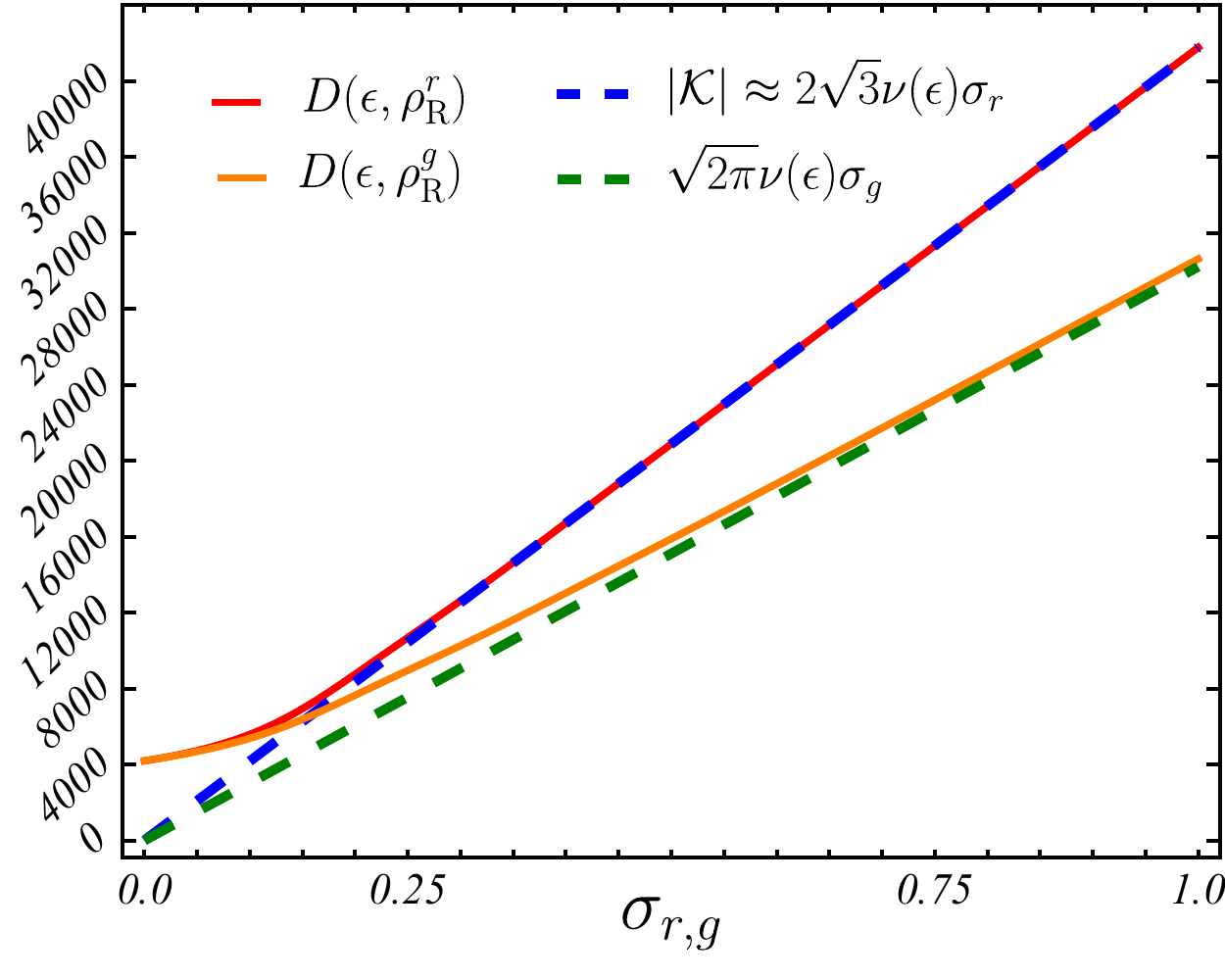}
    \caption{Analytical dimensionalities $D(\epsilon,\rho_R^r)$ and $D(\epsilon,\rho_R^g)$ for ensembles of  random states  with rectangular $\rho_\text{R}^{r}$ (red solid thin line) [see Eq.~\eqref{eq:neffrectangularprofile}] and Gaussian $\rho_\text{R}^{g}$ (orange solid thin line) [see Eq.~\eqref{eq:neffgaussianprofile}] energy profile, respectively. Both energy profiles are centered in the classical energy shell at $\epsilon=-0.5$ and their dimensionalities are plotted as a function of their respective energy standard deviations $\sigma_{r}$ and $\sigma_{g}$. Dashed lines represent the asymptotic value of both dimensionalities for large profile energy standard deviations. Blue dashed line for the rectangular energy profile [see Eq.~\eqref{eq:recLim}], and green dashed  line for the Gaussian energy profile [see Eq.~\eqref{eq:gaussLim}]. The system size is $j=100$.}
    \label{fig03}
\end{figure}

Figure~\ref{fig03} illustrates these findings by plotting the dimensionality $D(\epsilon, \rho_{\text{R}})$ for the two previous ensembles of  random states with rectangular and Gaussian energy profiles, $\rho_{\text{R}}^{r}$ and $\rho_{\text{R}}^{g}$, as a function of their respective energy standard deviations $\sigma_{r}$ and $\sigma_{g}$. In all cases we consider  energy profiles  centered around  $\epsilon=-0.5$. We can see that,  for large   $\sigma_{r}$ and $\sigma_{g}$, the dimensionality depends on the form of the  energy profile with $D(\epsilon,\rho_R)$ growing faster for the rectangular profile than for the Gaussian profile. However, when $\sigma_{r},\sigma_{g} \rightarrow 0$, the dimensionalities of both profiles attain a minimal value that is independent of the energy profile and is given by evaluating Eq.~\eqref{eq:effDim} in the classical energy shell at $\epsilon=-0.5$, that is, $\mathcal{D}_{\text{eff}}(\epsilon=-0.5) = 4201$.

\section{Dimensionality and Participation Ratio}
\label{sec:EffectiveDimensionPR}

In this section we discuss the dimensionality $D(\epsilon,\rho_{\text{R}})$ as defined in Eq.~\eqref{eq:generaleffdim} and its relation with the so-called participation ratio
\begin{equation}
    \label{eq:PR}
    P_R=\bigg(\sum_k |c_k|^4 \bigg)^{-1},
\end{equation}
with $c_k = \expval{\phi_k |\psi }$. This measure is  widely employed to estimate the number of states of a given orthonormal basis $\{ \ket{\phi_k} \}$ that conform an arbitrary state $\ket{\psi}$. 

We are interested in studying the relationship between the dimensionality of an ensemble of  random pure states $|\psi_{\text{R}}\rangle$ and the corresponding  participation ratio $P_{R}$ defined in the energy eigenbasis $\{|\varphi_{k}\rangle\}$. To this end, as in the previous section, we build  random states $|\psi_\text{R}^r\rangle$ with a rectangular energy profile $\rho_{\text{R}}^{r}$, delimited by the energy interval $[\epsilon_{i},\epsilon_{f}]$. The numbers between $k_i$ and $k_f$ enumerate the eigenstates $|\varphi_{k}\rangle$ with eigenenergies $\epsilon_{i} \leq \epsilon_{k} \leq \epsilon_{f}$.  For each $k$ in the set of indices $\mathcal{K}=\{k_i, k_i+1, k_i+2,  \dots, k_f\},$
let 
\begin{equation}
    \label{eq:rectangularrandomstatecoefficients}
    c_k^r=\begin{cases} z_k/\sqrt{ \sum_{l\in \mathcal{K}}|z_l|^2   }&\text{if } k\in \mathcal{K} \\ 0&\text{otherwise} \end{cases}
\end{equation}
be the coefficients of these  random pure states $|\psi_{\text{R}}^{r}\rangle$. These coefficients are built sampling random numbers $z_k$ from a distribution with mean $0$. For example, we could take $z_k$ from a real normal distribution with unitary standard deviation [Gaussian Orthogonal Ensemble (GOE)], or we could take $z_k$ from a complex normal distribution with unitary standard deviation [Gaussian Unitary Ensemble (GUE)]. 

Using Eqs.~\eqref{eq:generaleffdim} and~\eqref{eq:eigenstateseffdimEp}, and taking Eq.~\eqref{eq:rectangularprofile} for the rectangular energy profile $\rho_{\text{R}}^{r}(\epsilon)$, where $\epsilon=(\epsilon_i+\epsilon_f)/2$ is taken at the center of the energy window, the average dimensionality is calculated as 
\begin{equation}
    \label{eq:Neffrandomrectangularstate}
    D(\epsilon,\rho_{\text{R}}^{r})
    = \energyshellaveragelong*{\expval{ \frac{ \exp(\frac{-(\epsilon_k-\epsilon)^2}{2\sigma_{\bm x}^2})}{\nu(\epsilon_k)\sqrt{2\pi}\sigma_{\bm x}}}_{k\in \mathcal{K}}}{\bm x}{\epsilon} ^{-1},
\end{equation}
where $\expval{\,\cdots \,}_{k\in \mathcal{K}}\equiv \frac{1}{|\mathcal{K}|}\sum_{k\in \mathcal{K}}\,\cdots\,$ denotes the average over the indices of $\mathcal{K}$ and $|\mathcal{K}| \approx 2\sqrt{3}\nu(\epsilon)\sigma_r$ (see App.~\ref{app:EffectiveWidthDerivationRectangular} for details). 
From Eq.~\eqref{eq:recLim} we obtain the asymptotic behaviors
\begin{equation}
    \label{dim_average_limiting_cs}
    D(\epsilon,\rho_{\text{R}}^{r})
    \approx \begin{cases} |\mathcal{K}| &\text{if } \epsilon_f-\epsilon_i\gg \overline{\sigma}_{\text{c}}(\epsilon) \\ \sqrt{2\pi}\nu(\epsilon)\overline{\sigma}_{\text{c}}(\epsilon)&\text{if } \epsilon_f - \epsilon_i \ll \overline{\sigma}_{\text{c}}(\epsilon)  \end{cases} . 
\end{equation}
On the other hand, the participation ratio for the same ensemble of  random states $|\psi_{\text{R}}^r\rangle$ is given by
\begin{equation}
    \label{average_PR}
    \expval{P_R(\psi_{\text{R}}^{r})}_{\psi_\text{R}^r}=  \frac{\expval{|z_k|^4}_{\psi_\text{R}^r}}{\expval{|z_k|^2}_{\psi_\text{R}^r}^{2}} \abs{\mathcal{K}}.
\end{equation}
From Eqs.~\eqref{dim_average_limiting_cs}~and~\eqref{average_PR}, important differences between the dimensionality and the participation ratio can be seen. The dimensionality $D(\epsilon,\rho_{\text{R}}^{r})$ becomes  independent  of the random numbers  $z_k$ and  only depends on the properties of the energy shell and  the  energy profile. For wide energy windows, $D(\epsilon,\rho_{\text{R}}^{r})$ grows as $|\mathcal{K}|$ does, but  for narrow energy intervals it tends to  the value $\sqrt{2\pi}\nu(\epsilon)\overline{\sigma}_{\text{c}}(\epsilon)$. In contrast, $\langle P_{R}(\psi_{\text{R}}^{r}) \rangle$ is always  proportional to $\abs{\mathcal{K}}$ with  the proportionality   factor  given by   the details of the specific distribution from where the numbers $z_k$ are sampled. If the numbers $z_k$ are sampled from a real normal distribution (GOE), then this factor equals ${\expval{|z_k|^4}}/{\expval{|z_k|^2}^2}=1/3$, and if the numbers $z_k$ are sampled from a complex normal distribution (GUE), then ${\expval{|z_k|^4}}/{\expval{|z_k|^2}^2}=1/2$.

\begin{figure}
    \centering
    \includegraphics[width=0.95\columnwidth]{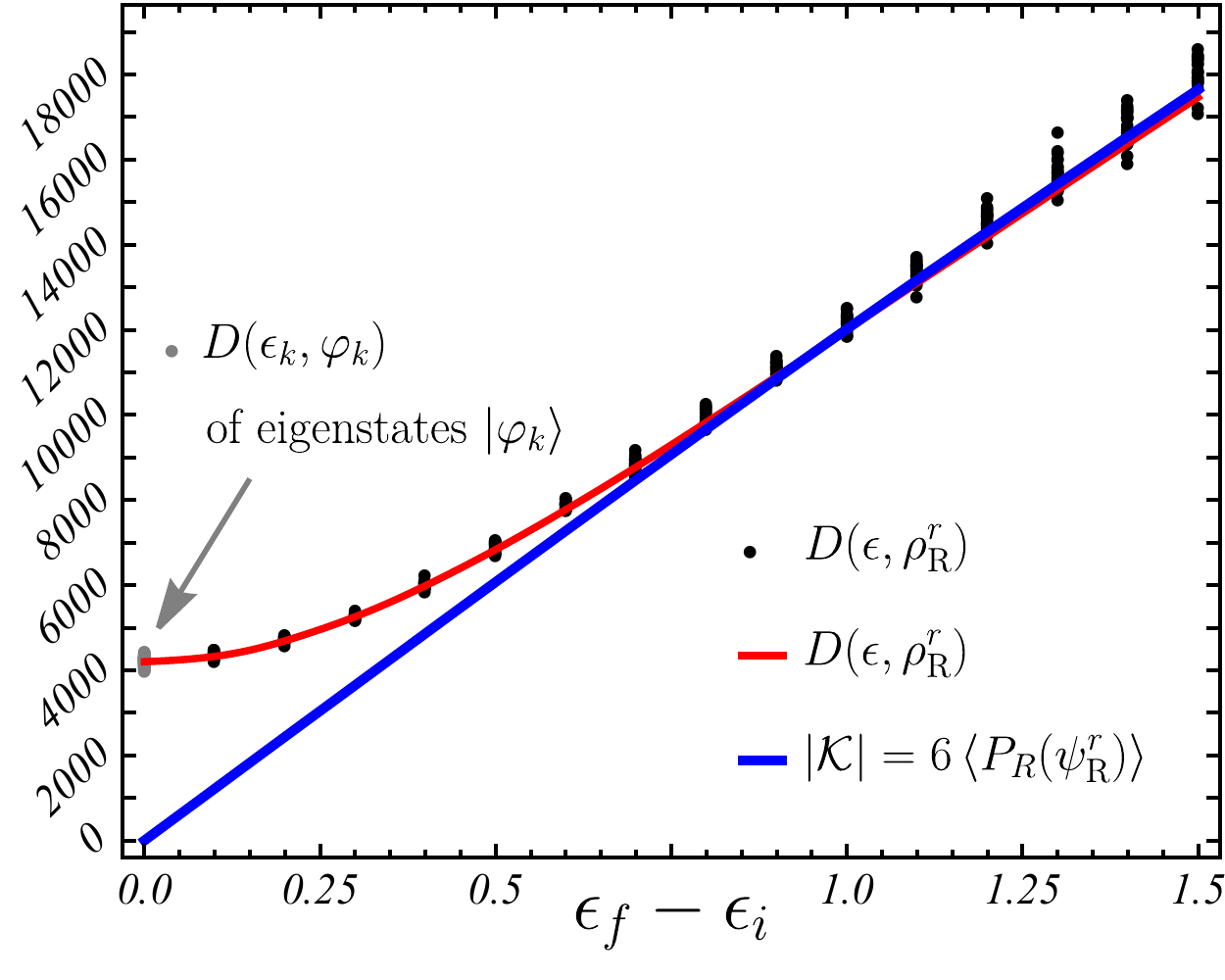}
    \caption{Dimensionality $D(\epsilon,\rho_{\text{R}}^{r})$ (black dots) for 20 single random states with rectangular energy profile $\ket{\psi_{\text{R}}^{r}}=\sum_{k\in\mathcal{K}'}c_{k}^{r}\ket{\varphi_k}$, whose coefficients $c_{k}^{r}$ are obtained according to Eq.~\eqref{eq:rectangularrandomstatecoefficients} [replacing $\mathcal{K}$ by $\mathcal{K}'$ given by Eq.~\eqref{eq:positiveparityindices}] with $z_k$ real-normally-distributed random numbers (GOE). The rectangular energy windows are centered in the classical energy shell at $\epsilon=-0.5$ and increase in standard deviation according to the horizontal axis. Analytical dimensionality $ D(\epsilon,\rho_{\text{R}}^{r})$ (red solid thin curve) for ensembles of the same random states as a function of the energy spacing $\epsilon_f-\epsilon_i=2\sqrt{3}\sigma_r$ [see Eq.~\eqref{eq:neffrectangularprofile}]. Number of eigenstates (blue solid thick line) within the rectangular energy window, which coincide with the average participation ratio for random states multiplied by a factor of 6, $|\mathcal{K}| = 6 \, \langle P_{R}(\psi_R^{r}) \rangle$ [see Eq.~\eqref{average_PR}]. Effective dimensions  $\mathcal{D}_{\text{eff}}(\epsilon_k)=D(\epsilon_k,\varphi_k)$ (gray dots) for 100 eigenstates $|\varphi_k\rangle$ around the classical energy shell at $\epsilon=-0.5$. The system size is $j=100$.}
    \label{fig04}
\end{figure}

Moreover, selecting only some of the eigenstates (like those inside some symmetry subspace of the system) and working only with a subset of $\mathcal{K}$, the participation ratio would evidently decrease in proportion to the number of eigenstates that were discarded. Instead, note that in Eq.~\eqref{eq:Neffrandomrectangularstate} there is no direct dependence on the size of $\mathcal{K}$. The dependence on $\mathcal{K}$  in Eq.~\eqref{eq:Neffrandomrectangularstate} is only indirect, through an average over its indices. One may sample only a subset of these indices and, assuming that it is a uniform sampling, the average would remain the same. This means that if one replaces $\mathcal{K}$ with a proper subset $\mathcal{K}'\subseteq \mathcal{K}$ where the indices are uniformly distributed, the dimensionality of the resulting random states is unchanged.

As an illustration, in Fig.~\ref{fig04} we show  the  dimensionality for random states $|\psi_{\text{R}}^{r}\rangle$ over rectangular windows of varying standard deviation in the chaotic region of the Dicke model. We build these states taking random numbers $z_k$ from a real standard normal distribution (GOE), and we only select eigenstates which have  positive parity, that is, we work with the indices in 
\begin{equation}
    \label{eq:positiveparityindices}
    \mathcal{K}'=\left\{k\in \mathcal{K}\mid \expval{\hat{\Pi}}{E_k}=1\right \},
\end{equation}
where the parity operator is defined as $\hat{\Pi}=\text{exp}[i\pi(\hat{a}^{\dagger}\hat{a}+\hat{J}_{z}+j\hat{1})]$. Because of these selections, the  average of participation ratio   is equal to $1/6$th of the total number of eigenstates inside the rectangular energy window $\abs{\mathcal{K}}$. A factor $1/3$ comes from the GOE sampling, and an additional factor $1/2$ comes from the fact that half of the eigenstates have positive parity, that is $\abs{\mathcal{K}'}=\abs{\mathcal{K}}/2$. It is clearly seen that, for a wide energy window, the dimensionality is equal to $\abs{\mathcal{K}}$, but, when the standard deviation of this window is small, it tends to the minimum value which is very similar to the effective dimension $\mathcal{D}_{\text{eff}}(\epsilon_k)$, where $\epsilon_k\approx\epsilon$ are  eigenvalues around the center of the energy window at $\epsilon=-0.5$.

Both the dimensionality $D$ and the participation ratio $P_{R}$ are measures of the number of states in a set required to describe an arbitrary state, nevertheless they have important differences as outlined above. All these differences ultimately stem from the fact that the dimensionality not only probes the dimension of the state, but also the dimension of the family of coherent states in the classical energy shell $\mathcal{M}(\epsilon)$. This property is precisely what allows us to extract an effective dimension for the classical energy shell out of the dimensionality of ensembles of random states.

\section{Conclusions}
\label{sec:Conclusions}

We have shown that in an unbounded phase space, we can assign an effective dimension $\mathcal{D}_{\text{eff}}(\epsilon)$ to the classical energy shell of coherent states satisfying $\langle \bm{x}|\hat{H}_{D}|\bm{x}\rangle/j=\epsilon$. We introduced  the  dimensionality of an ensemble of random  states with a given energy profile as  the inverse of the double average of their  Husimi function, $D(\epsilon,\rho_{\text{R}})=\langle\energyshellaverage{\mathcal{Q}_{\psi_{\text{R}}}}{\epsilon}\rangle_{\psi_{\text{R}}}^{-1}$. The first average is performed over the classical energy shell, and the second over the ensemble of random states. The effective dimension of the classical energy shell  was obtained by minimizing the dimensionality with respect to all possible energy profiles. It was shown that the harmonic mean of the coherent-state energy standard deviations in the classical energy shell $\overline{\sigma}_{\text{c}}(\epsilon)$ introduces a lower bound in the dimensionality of random states and that the minimum of the dimensionality is attained for random states having  an energy profile that is heavily peaked in a single energy.

An analytical expression for the effective dimension of classical energy shells was obtained in  the Dicke model as $\mathcal{D}_{\text{eff}}(\epsilon)=\sqrt{2\pi}\nu(\epsilon)\overline{\sigma}_{\text{c}}(\epsilon)$, where  $\nu(\epsilon)$ is the energy density of states. It is remarkable that the relevant mean is harmonic and not standard. These two types of means  are similar but do not exactly coincide. It was also shown that the  effective dimension of a classical shell at energy $\epsilon$ is very close  to the reciprocal average  of the Husimi function of the eigenstates over the eigenenergy shells close to $\epsilon$. Because of the non-zero coherent-state energy standard deviation, the effective dimension of a classical energy shell grows as $j^{3/2}$, where $j$ is the system size.
 
Finally, we compared the dimensionality of random  states
$D$ with the standard quantum participation ratio $P_{R}$ with respect to the energy eigenbasis, and several differences arose. They stem from the fact that $P_{R}$ depends only on the state $|\psi_{\text{R}}\rangle$, while $D$ also depends on the family of coherent states $|\bm{x}\rangle$ in a classical energy shell. Moreover, the dimensionality of a random state with a rectangular energy profile remains unchanged if some of the participating states are removed, provided that the standard deviation of the energy window does not change. This indicates that $D$ probes the dimensionality of the whole system inside an energy interval, and effectively depends only on the energy profile of the random states. On the other hand,  when the standard deviation of the energy window is large, the  dimensionality of random states becomes equal to the number of states in the energy window employed to build the random states, and proportional to the participation ratio in the eigenbasis.  

In finite systems, the scaling of the localization of random states is related to the dimension of the Hilbert space. The same is true for the Dicke model, with the dimension replaced by the effective dimension we present here. Our method is readily generalizable to be employed in other systems where a Husimi function can be built, such as billiards \cite{Lozej2022}, as the effective dimension is independent of the particular Hamiltonian. We believe that the study of the effective dimensions we proposed in this work could shed light to phenomena like quantum scarring and localization in other systems, both finite and infinite. For example, it has turned out to be particularly important for the study of  R\'enyi occupations of pure states~\cite{Pilatowsky2021Identification}.

\section*{ACKNOWLEDGMENTS}

We thank Lea F. Santos for fruitful discussions and helpful comments. We acknowledge the support of the Computation Center - ICN, in particular to Enrique Palacios, Luciano D\'iaz, and Eduardo Murrieta. SP-C, DV, and JGH acknowledge financial support from the DGAPA- UNAM project IN104020, and SL-H from the Mexican CONACyT project CB2015-01/255702.

\appendix

\section{Dimensionality of Random Pure States}
\label{app:EffectiveWidthDerivation}

Consider a random pure state $\ket{\psi_{\text{R}}}=\sum_k c_k \ket{\varphi_k}$. Assuming that the phases of 
$c_k$ are uniformly distributed, we may approximate~\footnote{For a sufficiently large sequence of random complex numbers $z_k=a_k e^{i\theta_k}$, where $\theta_i$ are sampled uniformly from $[0,2\pi)$, one has $\abs*{\sum_k z_k}^{2}\approx \sum_k a_k^2,$ because the total distance $R=\abs{\sum_k z_k}$ is distributed according to a  Rayleigh distribution $P(R)={2R}\exp(-{R^2}/{a^2})/{a^2} \quad (a^2=\sum_k a_k^2)$
whose second moment is  $a^{2}$ (see Ref.~\cite{Hughes1995Book} p. 71, Eq. 2.77).} the  average of the Husimi function over a classical energy shell as
\begin{align}
    \energyshellaverage{\mathcal{Q}_{\psi_{\text{R}}}}{\epsilon}
    &=\energyshellaveragelong*{\abs{\braket{\bm x}{\psi_{\text{R}}}}^2}{\bm x}{\epsilon} \nonumber \\
    &=
   \energyshellaveragelong*{\Big |{\sum_k C_k(\bm x) c_k\Big |}^2}{\bm x}{\epsilon} \nonumber
   \\
   &\approx \left\langle \sum_k \abs{C_k(\bm x) c_k}^2\right\rangle_{\bm x\in \mathcal{M}(\epsilon)}\nonumber \\
   &= \sum_k \abs{ c_k}^2\energyshellaveragelong*{\abs{C_k({\bm x})}^2}{\bm x}{\epsilon}\nonumber\\
   &=\sum_k |c_k|^2  \energyshellaverage{\mathcal{Q}_{\varphi_k}}{\epsilon},
   \label{eq:appa1}
\end{align}
where $C_k(\bm x)=\braket{\bm x}{\varphi_k}$. When the previous result is substituted in Eq.~\eqref{eq:dimendef}, it leads to the expression for the dimensionality given by Eq.~\eqref{eq:dimenwithcomp}.


To further simplify the  dimensionality $D(\epsilon, \rho_{\text{R}})$, we   consider  averages over the  ensemble of random states $|\psi_{\text{R}}\rangle$ with energy profile $\rho_{\text{R}}(\epsilon)$. Inserting  Eq.~\eqref{eq:appa1} into Eq.~\eqref{eq:dimendef} gives
\begin{align}
    \label{eq:app2}
    D(\epsilon, \rho_{\text{R}})\equiv\left \langle \energyshellaverage*{\mathcal{Q}_{\psi_{\text{R}}}}{\epsilon} \right\rangle_{\psi_{\text{R}}}^{-1}=\left(\sum_k \langle|c_k|^2\rangle_{\psi_{\text{R}}}  \energyshellaverage{\mathcal{Q}_{\varphi_k}}{\epsilon}\right)^{-1}.
\end{align}

The ensemble average of the components can be calculated as follows. From Eq.~\eqref{eq:compon} we obtain
\begin{equation}
    \label{eq:composq}
    \langle |c_k|^2\rangle_{\psi_{\text{R}}}= \left\langle \frac{r_k}{M}\right \rangle_{\psi_{\text{R}}} \frac{\rho_{\text{R}}(\epsilon_k)}{\nu(\epsilon_k)},
\end{equation}
where $M$ is a normalization constant given by $M=\sum_l r_l \, \rho_{\text{R}}(\epsilon_l) \, / \, \nu(\epsilon_l)$. Therefore, the ensemble average of the ratio $r_{k}/M$ is
\begin{align}
    \left\langle \frac{r_k}{M}\right \rangle_{\psi_{\text{R}}} & \approx \frac{\langle r_k\rangle_{\psi_{\text{R}}}}{\sum_l \langle r_l\rangle_{\psi_{\text{R}}}\rho_{\text{R}}(\epsilon_l)/\nu(\epsilon_l)} \nonumber \\
    & \approx  \frac{1}{\sum_l \rho_{\text{R}}(\epsilon_l)/\nu(\epsilon_l)},
\end{align}
where the average  $\langle r_l\rangle_{\psi_{\text{R}}}$ is actually independent of the index $l$ and cancels the average in the numerator. Now,  approximating the sum by an integral $\sum_l \,\bullet\ \rightarrow \int \dd \epsilon' \, \nu(\epsilon')\,\bullet\,$  yields
\begin{equation}
    \left\langle \frac{r_k}{M}\right \rangle_{\psi_{\text{R}}}\approx 1,
\end{equation}
where we used the normalization $\int \dif \epsilon \, \rho_{\text{R}}(\epsilon) = 1$. By substituting this result in \eqref{eq:composq} and then in \eqref{eq:app2} we obtain Eq.~\eqref{eq:generaleffdim}.

\section{Average of the Husimi Function of Eigenstates over Classical Energy Shells}
\label{app:AveHusEigen}
To compute the average $\energyshellaveragelong{\abs{C_k({\bm x})}^2}{\bm x}{\epsilon}=
\energyshellaverage{\mathcal{Q}_{\varphi_k}}{\epsilon}$, we use the fact that most coherent states of the Dicke model have a random-like  structure~\cite{Lerma2018, Villasenor2020}
\begin{equation*}
    \abs{C_k(\bm x)}^2=\frac{r_k(\bm x) G_{\bm x}(\epsilon_k)}{\nu(\epsilon_k)M(\bm x)},
\end{equation*}
where $\epsilon_k=E_k/j$, $\nu(\epsilon)$ is the density of states, $r_k(\bm x)$ is a random number sampled from a positive distribution, and $G_{\bm x}(\epsilon)$ is the normalized continuous energy profile of the coherent states. For the high-energy regime studied here,  $G_{\bm x}$ are Gaussian functions centered at $h_\text{cl}(\bm x)=\epsilon$ with energy standard deviation $\sigma_{\bm x}$ (see App.~\ref{app:EnergyWidthDerivation}), that is~\cite{Villasenor2020}
$$
    G_{\bm x}(\epsilon_k)=\frac{\exp(
    \frac{-(\epsilon_k-\epsilon)^2}{2\sigma_{\bm x}^2})}{\sqrt{2\pi}\sigma_{\bm x}}.
$$ 
The number $M(\bm x)=\sum_l r_l(\bm x) G_{\bm x}(\epsilon_l)/\nu(\epsilon_l)$ ensures normalization. Thus,
\begin{align*}
  \energyshellaverage{\mathcal{Q}_{\varphi_k}}{\epsilon}&=  \energyshellaveragelong*{\abs{C_k({\bm x})}^2}{\bm x}{\epsilon}\\
  &= \energyshellaveragelong*{\frac{r_k(\bm x) G_{\bm x}(\epsilon_k)}{\nu(\epsilon_k)M(\bm x)}}{\bm x}{\epsilon}\\&\approx \frac{\energyshellaveragelong*{r_k(\bm x) G_{\bm x}(\epsilon_k)}{\bm x}{\epsilon}}{\nu(\epsilon_k)\energyshellaveragelong*{M(\bm x)}{\bm x}{\epsilon}}\\&=\frac{\energyshellaverage*{r_k}{\epsilon}\energyshellaveragelong*{ G_{\bm x}(\epsilon_k)}{\bm x}{\epsilon}}{\nu(\epsilon_k)\sum_l \energyshellaverage*{r_l}{\epsilon}\energyshellaveragelong*{ G_{\bm x}(\epsilon_l)}{\bm x}{\epsilon}/\nu(\epsilon_l)}.
\end{align*}
The average $\energyshellaverage*{r_l}{\epsilon}$ actually does not depend on $l$, so it cancels $\energyshellaverage*{r_k}{\epsilon}$. Furthermore, we may approximate the sum $\sum_l \,\bullet\,$ by the integral $\int \dd \epsilon' \, \nu(\epsilon')\,\bullet\,$, so due to normalization
\begin{align*}
    \sum_l \frac{\energyshellaveragelong*{G_{\bm x}(\epsilon_l)}{\bm x}{\epsilon}}{\nu(\epsilon_l)} &=\int \dd \epsilon' \energyshellaveragelong*{G_{\bm x}(\epsilon')}{\bm x}{\epsilon} \\
    &=\energyshellaveragelong*{\int \dd \epsilon'\, G_{\bm x}(\epsilon')}{\bm x}{\epsilon}\\
    &=\expval{1}_{\bm x\in \mathcal{M}(\epsilon)}\\
    &=1.
\end{align*}
Thus, the average
\begin{equation*}
    \energyshellaverage{\mathcal{Q}_{\varphi_k}}{\epsilon}=   \energyshellaveragelong*{\abs{C_k({\bm x})}^2}{\bm x}{\epsilon}\approx\frac{\energyshellaveragelong{G_{\bm x}(\epsilon_k)}{\bm x}{\epsilon}}{\nu(\epsilon_k)},
\end{equation*}
is independent of the distribution of $r_k$. Finally, using the fact that $G_{\bm x}$ is a Gaussian we obtain Eq.~\eqref{eq:eigenstateseffdimEp}.



\section{Energy Standard Deviation of Coherent States}
\label{app:EnergyWidthDerivation}

The variance of the quantum Hamiltonian $\hat{H}_{D}$ [see Eq.~\eqref{eqn:qua_hamiltonian}] under the tensor product of Glauber-Bloch coherent states $|\bm{x}\rangle=|q,p\rangle\otimes|Q,P\rangle$ [see Eq.~\eqref{eq:coherentstates}] is given by~\cite{Lerma2018}
\begin{align}
    \label{eq:CSenergywidth}
    \sigma_{\bm x}^2 & = \frac{1}{j^{2}}\left[\langle\bm{x}|\hat{H}_{D}^{2}|\bm{x}\rangle-\langle\bm{x}|\hat{H}_{D}|\bm{x}\rangle^{2}\right] \nonumber \\
    & = \frac{1}{j^2}\left[\Omega_{1}(\bm{x})+\Omega_{2}(\bm{x})\right].
\end{align}
The terms $\Omega_{1}(\bm{x})$ and $\Omega_{2}(\bm{x})$ are given explicitly by
\begin{align}
    \Omega_{1}(\bm{x}) = & j\left\{ \frac{\omega^{2}}{2}\left(q^{2}+p^{2}\right) + \frac{\omega_{0}^{2}}{2}\left(Q^{2}+P^{2}\right)A^{2}(Q,P) \right. + \nonumber \\
    & 2\gamma^{2}\left[\frac{q^{2}}{\gamma^{2}}\Omega_{2}(\bm{x}) + Q^{2}A^{2}(Q,P)\right] + \nonumber \\
    & \left. 2\gamma qQ\left[\omega+\omega_{0}\left(1-\frac{Q^{2}+P^{2}}{2}\right)\right]A(Q,P) \right\}, \nonumber \\
    \Omega_{2}(\bm{x}) = & \gamma^{2}\left[P^{2}A^{2}(Q,P)+\left(1-\frac{Q^{2}+P^{2}}{2}\right)^{2}\right],
\end{align}
with $A(Q,P)=\sqrt{1-\frac{Q^{2}+P^{2}}{4}}$.

\section{Dimensionalities for Ensembles with Rectangular and Gaussian Profiles}
\subsection{Dimensionality for a Rectangular Energy Profile}
\label{app:EffectiveWidthDerivationRectangular}

From 
Eq.~\eqref{eq:generaleffdim} and Eq.~\eqref{eq:eigenstateseffdimEp} for $\energyshellaverage{\mathcal{Q}_{\varphi_k}}{\epsilon}$, and  considering   the rectangular energy profile $\rho_{\text{R}}^{r}(\epsilon)$ of  Eq.~\eqref{eq:rectangularprofile}, we obtain
\begin{align}
    D(\epsilon,\rho_{\text{R}}^{r}) 
    & = \energyshellaveragelong*{\sum_{k\in\mathcal{K}} \frac{ \exp(\frac{-(\epsilon_k-\epsilon)^2}{2\sigma_{\bm x}^2})}{2\sqrt{3}\nu^{2}(\epsilon_k)\sigma_{r}\sqrt{2\pi}\sigma_{\bm x}}}{\bm x}{\epsilon} ^{-1} \nonumber \\
    &= \energyshellaveragelong*{\frac{1}{|\mathcal{K}|}\sum_{k\in \mathcal{K}} \frac{ \exp(\frac{-(\epsilon_k-\epsilon)^2}{2\sigma_{\bm x}^2})}{\nu(\epsilon_k)\sqrt{2\pi}\sigma_{\bm x}}}{\bm x}{\epsilon} ^{-1}, 
\end{align}
where in the second line we used $|\mathcal{K}| \approx 2\sqrt{3}\nu(\epsilon)\sigma_r$. Evaluating the density of states at the center of the rectangular profile $\nu(\epsilon)$, we obtain Eq.~\eqref{eq:Neffrandomrectangularstate}. On the other hand, if we approximate the sum $\sum_{k\in \mathcal{K}}\, \bullet$ by the integral $\int_{\epsilon_i}^{\epsilon_f} \dif \epsilon'\nu(\epsilon')\,\, \bullet\,$, we obtain the following simplified expression for the dimensionality
\begin{align}
    D(\epsilon,\rho_{\text{R}}^{r}) 
    & = \energyshellaveragelong*{\int_{\epsilon-\sqrt{3}\sigma_{r}}^{\epsilon+\sqrt{3}\sigma_{r}} \dif \epsilon' \frac{ \exp(\frac{-(\epsilon'-\epsilon)^2}{2\sigma_{\bm x}^2})}{2\sqrt{6\pi}\nu(\epsilon')\sigma_{r}\sigma_{\bm x}}}{\bm x}{\epsilon} ^{-1}, 
\end{align}
which gives Eq.~\eqref{eq:neffrectangularprofile} when the density of states is evaluated in the center of the rectangular profile $\nu(\epsilon)$.

It is straightforward to analyze the limiting behaviors of 
Eq.~\eqref{eq:neffrectangularprofile}. For $\sigma_r\gg \overline{\sigma}_{\text{c}}$ the error function is $\text{erf}(x)\approx 1$ and the average over the classical energy shell is also $1$. 
In the opposite limit, if $\sigma_r\ll \overline{\sigma}_{\text{c}}$, the error function is approximated by  $\text{erf}(x)\approx 2x/\sqrt{\pi}$. These two limits lead to Eq.~\eqref{eq:recLim}.

\subsection{Dimensionality for a Gaussian Energy Profile}
\label{app:EffectiveWidthDerivationGauss}

A similar procedure can be applied to the Gaussian energy profile, $\rho_{\text{R}}^g(\epsilon)$ given by Eq.~\eqref{eq:gaussianprofile}. From  Eqs.~\eqref{eq:generaleffdim} and~\eqref{eq:eigenstateseffdimEp},  we obtain
\begin{align}
    D(\epsilon,\rho_{\text{R}}^{g}) 
    & = \energyshellaveragelong*{\int_{-\infty}^{+\infty}  \!\!\!\!\dif \epsilon'\frac{\exp[\frac{-(\epsilon'-\epsilon)^2}{2}\left( \sigma_g^{-2}+\sigma_{\bm{x}}^{-2}\right)]}{2\pi \nu(\epsilon') \sigma_g\sigma_{\bm{x}}}  }{\bm x}{\epsilon}^{-1}. 
\end{align} 
By evaluating  the density of states at the center of the Gaussian  profile $\nu(\epsilon)$  and performing the Gaussian integral we obtain Eq.~\eqref{eq:neffgaussianprofile}. Moreover, the limiting behaviors of Eq.~\eqref{eq:neffgaussianprofile} can be obtained in a similar way to the rectangular case. For $\sigma_g \gg \overline{\sigma}_{\text{c}}$ we use the approximation $(1+x^{2})^{-1/2}\approx 1$, and for $\sigma_g \ll \overline{\sigma}_{\text{c}}$ we use $(1+x^{2})^{-1/2}\approx x^{-1}$. These two limiting behaviors lead to Eq.~\eqref{eq:gaussLim}.

\bibliography{bibliography}

\end{document}